\documentclass[graybox,natbib]{svmult}
\usepackage{placeins}         % to fix floating figures in corresponding section
\usepackage{natbib}
\usepackage{subfig}
\usepackage{mathptmx}       % selects Times Roman as basic font
\usepackage{helvet}         % selects Helvetica as sans-serif font
\usepackage{courier}        % selects Courier as typewriter font
\usepackage{makeidx}         % allows index generation
\usepackage[pdftex]{graphicx}  % standard LaTeX graphics tool
\usepackage{multicol}        % used for the two-column index
\usepackage[bottom]{footmisc}% places footnotes at page bottom
\usepackage{subfig}
\makeindex             % used for the subject index
\begin{document}

\title*{Numerical simulations of the Kelvin-Helmholtz instability with the Gadget-2 SPH code}

\author{{\bf Ruslan F. Gabbasov, Jaime Klapp-Escribano, Joel Su\'arez-Cansino
and Leonardo Di G. Sigalotti}}

\institute{R.F. Gabbasov \at Universidad Aut\'onoma del Estado de
Hidalgo, Instituto de Ciencias B\'asicas e Ingenier\'\i{}a, Ciudad
Universitaria, Km 4.5 Carretera Pachuca -- Tulancingo, 
Mineral de la Reforma, 42184, Hidalgo, M\'exico.
\email{ruslan.gabb@gmail.com} \and J. Klapp-Escribano \at Instituto
Nacional de Investigaciones Nucleares, Carretera M\'exico-Toluca Km
36.5, La Marqueza 52750, Estado de M\'exico, M\'exico.
\email{jaime.klapp@inin.gob.mx} \and J. Klapp-Escribano \at
Departamento de Matem\'aticas, Cinvestav del IPN, M\'exico 07360
D.F., M\'exico. \email{jaime.klapp@hotmail.com} \and J.
Su\'arez-Cansino \at Universidad Aut\'onoma del Estado de Hidalgo,
Instituto de Ciencias B\'asicas e Ingener\'\i{}a,
 Ciudad Universitaria, Km 4.5 Carretera Pachuca -- Tulancingo, 
Mineral de la Reforma, 42184, Hidalgo, M\'exico.
\email{jsuarez@uaeh.edu.mx}
\and Leonardo Di G. Sigalotti \at Centro de F\'\i{}sica, Instituto 
Venezolano de Investigaciones Científicas, IVIC,
Apartado Postal 20632, Caracas 1020-A, Venezuela.
\email{leonardo.sigalotti@gmail.com}
}

\authorrunning{Gabbasov et al.}

%\thispagestyle{empty}
%\maketitle
%\thispagestyle{empty}
%\setcounter{page}{1}

\thispagestyle{empty} \maketitle \thispagestyle{empty}

\abstract{The method of Smoothed Particle Hydrodynamics (SPH) has been
widely studied and implemented for a large variety of problems,
ranging from astrophysics to fluid dynamics and elasticity problems
in solids. However, the method is known to have several deficiencies
and discrepancies in comparison with traditional mesh-based codes.
In particular, there has been a discussion about its ability to
reproduce the Kelvin-Helmholtz Instability in shearing flows.
Several authors reported that they were able to reproduce correctly
the instability by introducing some improvements to the algorithm.
In this contribution, we compare the results of the \citet{Read10}
implementation of the SPH algorithm with the original Gadget-2
N-body/SPH code.}

\section{Introduction}
%\label{s:intro}
The Kelvin-Helmholtz Instability (KHI) is the instability that
appears at the interface between two shearing fluid flows of
different densities. Many experimental and numerical results have
been published where such instability is reproduced. In particular,
in astrophysics such instabilities may be responsible for many
phenomena observed in regions of high gas density contrast
\citep{Murray93}. The necessity for subsonic velocities for the KHI
survival was first explored in the context of astrophysical flows by
\citet{Vietri97}. As discussed by these authors, the growth rate of
the KHI is smaller than the gas speed of sound. The correct modeling
of KHI is essential since the vorticity and shear flows appear in
diverse hydrodynamic processes,
such as for example, the onset of turbulence.

The Smoothed Particle Hydrodynamics method (SPH) is a Lagrangian
meshless particle method used for simulation of fluid transport.
Since its original formulation the SPH method has been constantly
improved. In recent years a search for possible differences between
grid-based and particle-based methods has been widely discussed. For
instance, \citet{Agertz07} compare the results
for several
hydrodynamic tests obtained with SPH and grid-based methods. They
found a striking difference between the results. In particular, the
inability to reproduce the vorticity rolls in the shearing flow was
claimed to be a deficiency of the SPH method. \citet{Price08} showed
that introducing an artificial thermal conductivity term into the
standard SPH, in order to smooth the discontinuities in the thermal
energy, allowed for similar results to those obtained by
\citet{Agertz07} using grid-based simulations.

On the other hand, several authors suggested that the artificial
conductivity is not the main factor responsible for suppression of
the instability. It is well known that the choice of the smoothing
kernel also affects the results. For example, the standard cubic
spline kernel tends to suffer from the clumping instability (known
also as pairing instability) for large number of neighbors,
introducing errors and lowering the resolution. \citet{Read10} showed
that kernels that produce a constant force term in the center
prevent the clumping. Several alternative kernel shapes were
proposed to remedy this problem (see for example, \citet{Dehnen12}).
\citet{Hubber13} performed a comparison of KHI simulations obtained
with SPH and with the AMR Eulerian code {\em Pencil}. They concluded
that convergence between SPH and grid codes may be obtained if
higher order kernels (i.e., quintic) that support larger numbers of
neighbors are used. \citet{Cha10} showed that inaccurate density
gradients that are obtained with the standard SPH formulation are
responsible for the suppression of instabilities and can be
alleviated by using a Godunov formulation of SPH. These works
concluded that the standard SPH formulation is unable to reproduce
the KHI and that some improvements must be implemented.

Some of the causes that suppress the KHI are the following: when
pressure discontinuities appear as a result of the lack of entropy
mixing on the kernel scale, when large errors are introduced in the
momentum equation due to a finite number of neighbor particles, due to
the pairing instability, or due to contact discontinuities.

In this work, we compare the results for the KHI as obtained using the standard
Gadget-2 SPH with the formulation proposed by \citet{Read10}.

\section{Initial conditions}
As the initial conditions for the first part of the KHI tests (A), we
use the example already included in the OSPH code \citep{Read10}.
This is done with the aim of providing direct comparison of our simulations with
previously published results. In the second part (B), we use the
so-called ``well-posed'' initial conditions described in
\citet{McNally12} (see also \citet{Robertson10}). The latter is
regularized by smoothing the density and velocity on the interfaces
in order to prevent abrupt pressure jumps. Such situations are
frequently found for example in simulations of real astrophysical
systems. In addition, the two sets of initial conditions differ by
the type of density sampling, which is done by the spatial distribution
of the particles (set A) or by varying the particle masses (set B). Set
C differs from set B only in the magnitude of the densities and
velocities.

For brevity, we describe only the initial conditions for the
simulation sets B and C, which consist of a 3D slab of size $1.0\times
1.0\times 0.0325$, with the central part having a density
$\rho_2=2\rho_1$ and moving with the velocity ${\mbox v}_2=-u$,
while the upper and lower sections have a density $\rho_1$ and move
with the velocity ${\mbox v}_1= u$ along the $x$-axis. A small
velocity perturbation is added, ${\mbox v}_y=\delta {\mbox v}_y
\sin(2\pi x/\lambda)$, with $\lambda=0.5$. The gas has a constant
pressure $P=2.5$ everywhere and satisfies an ideal gas equation of
state $P=(\gamma-1)U\rho$, with $\gamma=5/3$. The system of units is
such that length, time and mass are equal to unity. For the
simulation set B, we use $\rho_1=1.0$, $u=0.5$ and $\delta {\mbox
v}_y=0.01$, while for the set C we define $\rho_1=32.0$, $u=0.1$,
$\delta {\mbox v}_y=0.01$ and $\delta {\mbox v}_y=0.002$. The system
is sampled with an equally spaced cubic grid of $256\times 256\times
8$ particles. The parameters are chosen with the purpose of
staying in the subsonic regime, which guarantees the KHI formation.

The characteristic onset time of the KHI in the linear regime is
given by:
\begin{equation}
\tau = \frac{(\rho_1+\rho_2)\lambda}{\sqrt{\rho_1\rho_2}\ |\vec{v}_2-\vec{v}_1|}\; .
\end{equation}
For the above initial parameters the characteristic times are $\tau_A\approx 3.4$,
$\tau_B \approx 1.06$ and $\tau_C \approx 5.3$, and the total
simulation times were $4$, $2.1$, and $8$ respectively. Note that different units
are used for set A.

We use the Gadget-2 code \citep{Springel05} and its modification --
OSPH -- the Optimized SPH introduced by \citet{Read10}. In both
codes the flags \verb|-DPERIODIC|, \verb|-DNOGRAVITY| and
\verb|-DLONG| were activated, and in OSPH the recommended flags were
also switched on (\verb|-DOSPHM|, \verb|-DOSPHRT|, and
\verb|-DOSPHHOCT|). The artificial viscosity was set to $\alpha=0.8$
in all simulations.
\section{Results}
\label{res} 
\begin{figure*}[ht]
\centering
\subfloat{\includegraphics[width=0.5\textwidth]{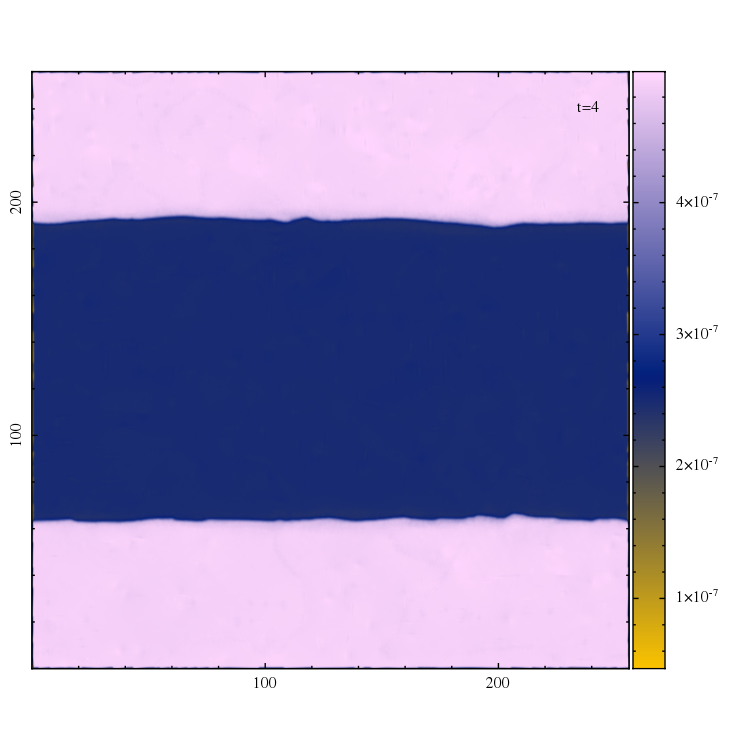} }
\subfloat{\includegraphics[width=0.5\textwidth]{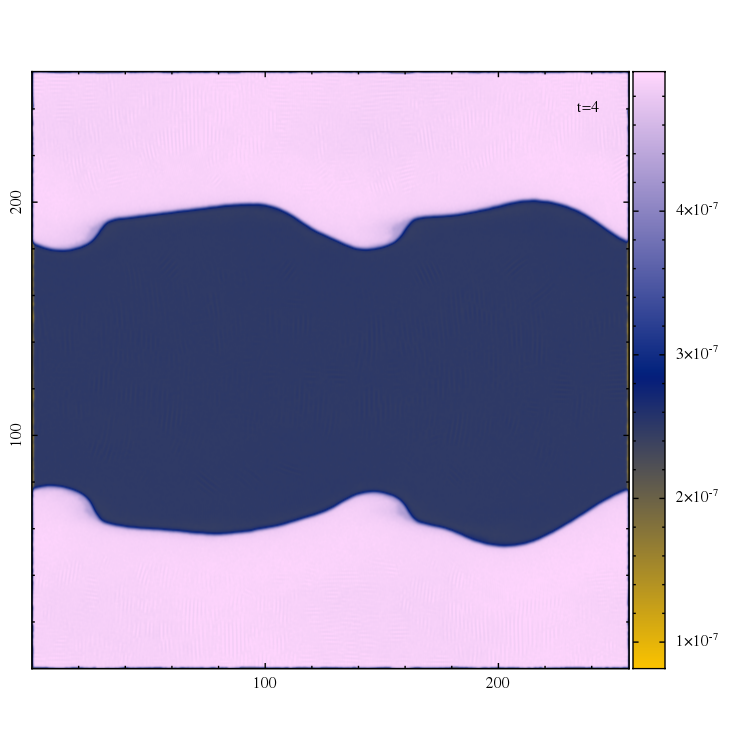} }\\
\vspace{-15pt}
\subfloat{\includegraphics[width=0.5\textwidth]{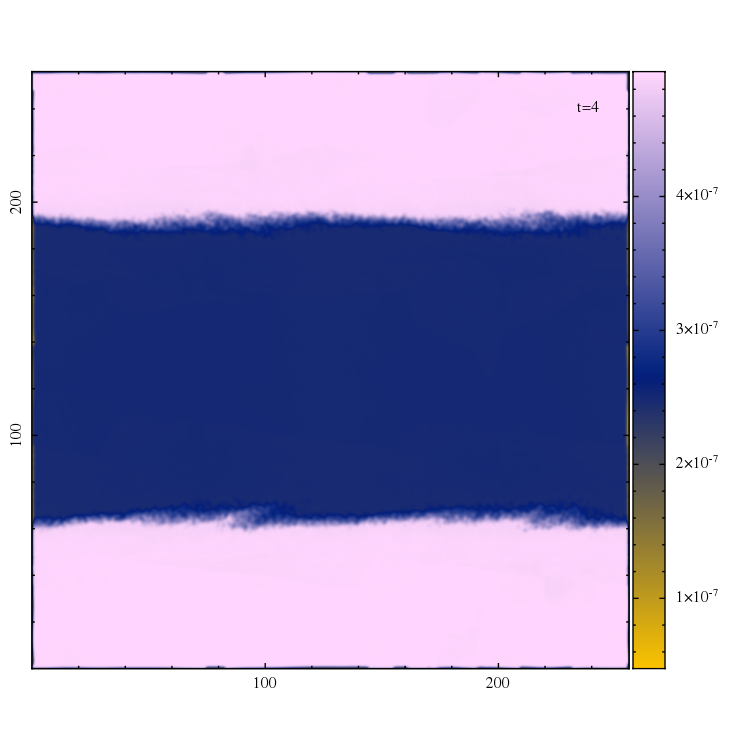} }
\subfloat{\includegraphics[width=0.5\textwidth]{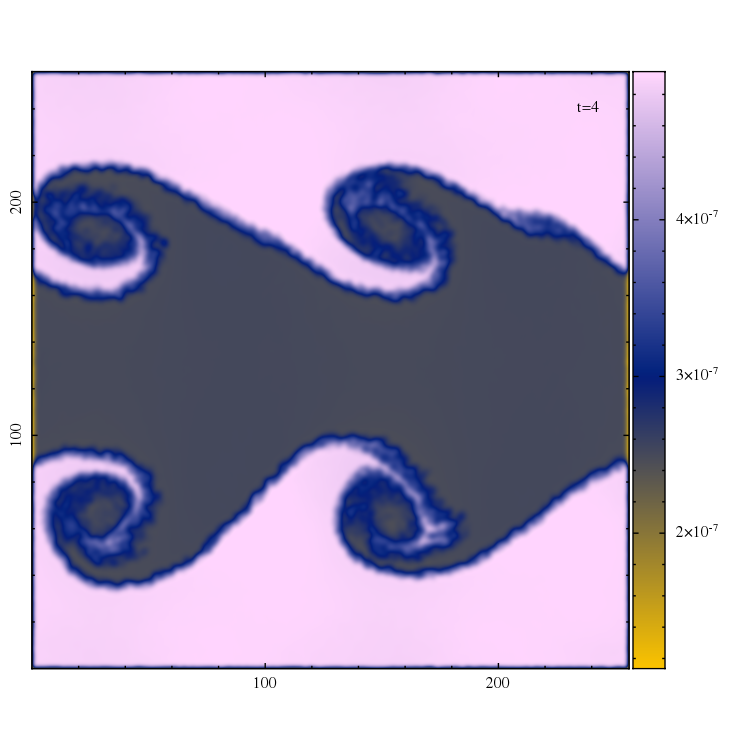} }
\vspace{-15pt}
\caption{Projected density plots for the simulations
of set A showing the perturbations at $1.2\tau_A$. From left to
right and top to bottom: AGADG33, AGADG64, AOSPH96, AOSPH442.}
\label{A}
\end{figure*}

We performed a number of tests using different combinations of
the initial conditions and code parameters. In Table \ref{tab1} we
summarize some of them. The first column is the set of models coded
by the first letter of the name, which are A, B or C, the second
column gives the number of neighbors, and the third column shows the
code employed for the simulation. Figure \ref{A} (set A), Fig. \ref{B} (set B) and Fig.
\ref{C} (set C) summarize the results of using different initial
conditions and different numbers of neighbor particles.
\begin{table*}
\centering \caption{Summary of the KHI simulations}
\label{tab:1}       % Give a unique label
\begin{tabular}{p{2cm}p{2.4cm}p{2.9cm}}
\hline\noalign{\smallskip}
Model & N$_{ngb}$ & Code  \\
\noalign{\smallskip}\svhline\noalign{\smallskip}
AGADG33 & 33 & Gadget-2 \\
AGADG64 & 64$^a$ & Gadget-2 \\
AOSPH96 & 96  &  OSPH\\
AOSPH442 & 442  &  OSPH\\
\noalign{\smallskip}
BGADG64 & 64$^a$  & Gadget-2\\
BOSPH64 & 64  &  OSPH\\
BOSPH96 & 96  &  OSPH\\
BOSPH442 & 442  &  OSPH\\
\noalign{\smallskip}
CGADG64L & 64  &  Gadget-2\\
CGADG64S & 64  &  Gadget-2\\
COSPH96L & 96  &  OSPH\\
COSPH96S & 96  &  OSPH\\
\noalign{\smallskip}\hline\noalign{\smallskip}
\end{tabular}\\
$^a$ The runs with a higher number of neighbors with the standard Gadget-2 were not\\
completed because a maximum number of tree-nodes was reached due
to the pairing instability. \label{tab1}
\end{table*}

From Fig. \ref{A}, we observe that model AOSPH442 is the only run
that produces well distinguished rolls, while using either a smaller number
of neighbors or the Gadget-2 code leads only to mild perturbations.
This fact confirms the results of \citet{Read10}. A notable feature
is that in model AOSPH96 with a smaller number of neighbors the perturbations
are completely damped. On the other hand, run AGADG64, which uses a nearly
optimum number of neighbors for the 3D cubic spline kernel, shows
pronounced undulations, although without any rolls. Using their new
SPHS code, which implements a dissipation switch, \citet{Read12}
obtained KHI in both single mass and multimass particle models.
Their initial conditions for single mass particles are identical to
those used in set A. Note that in the set A a Mach number for the
low density layer is $M\approx 0.11$.
\begin{figure*}[ht]
\centering
 \subfloat{ \includegraphics[width=0.5\textwidth]{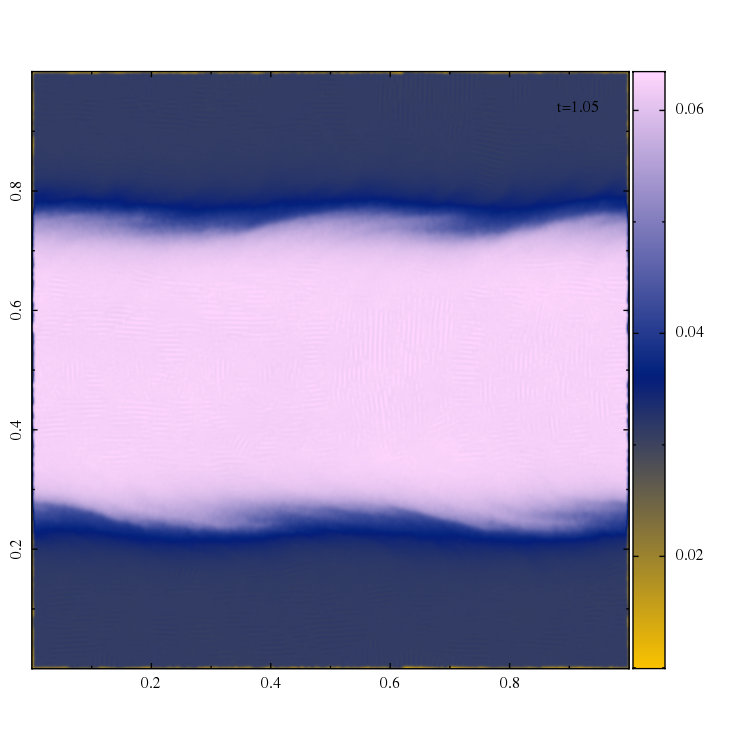} }
 \subfloat{ \includegraphics[width=0.5\textwidth]{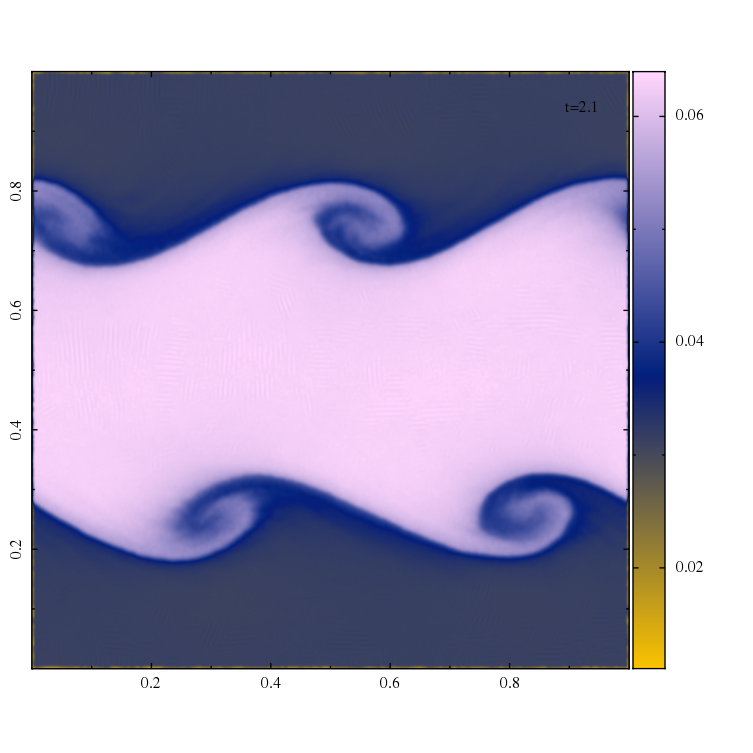} }\\
\vspace{-15pt}
 \subfloat{ \includegraphics[width=0.5\textwidth]{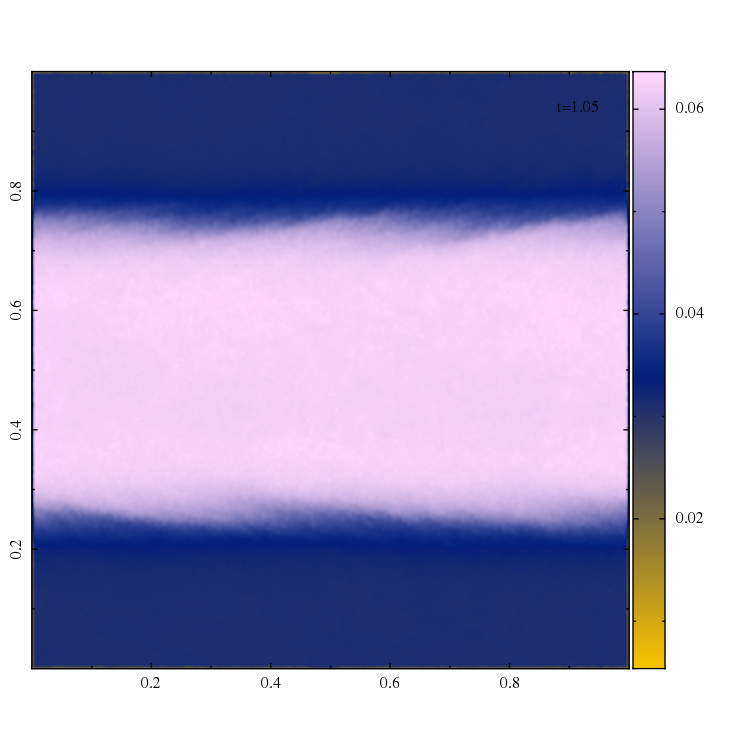} }
 \subfloat{ \includegraphics[width=0.5\textwidth]{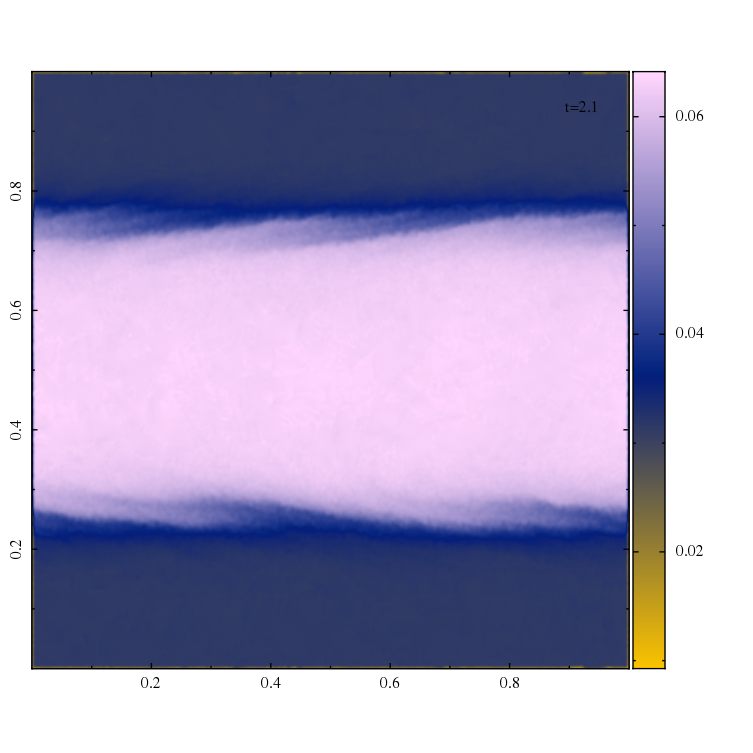} }\\
\vspace{-15pt}
\caption{Projected density plots for the simulations of set B. Each
column shows the perturbations at $\tau_B$ (left) and $2\tau_B$
(right). From top to bottom: BGADG64 and BOSPH64.} \label{B}
\end{figure*}

\begin{figure*}[ht]
\ContinuedFloat
\centering
 \subfloat{ \includegraphics[width=0.5\textwidth]{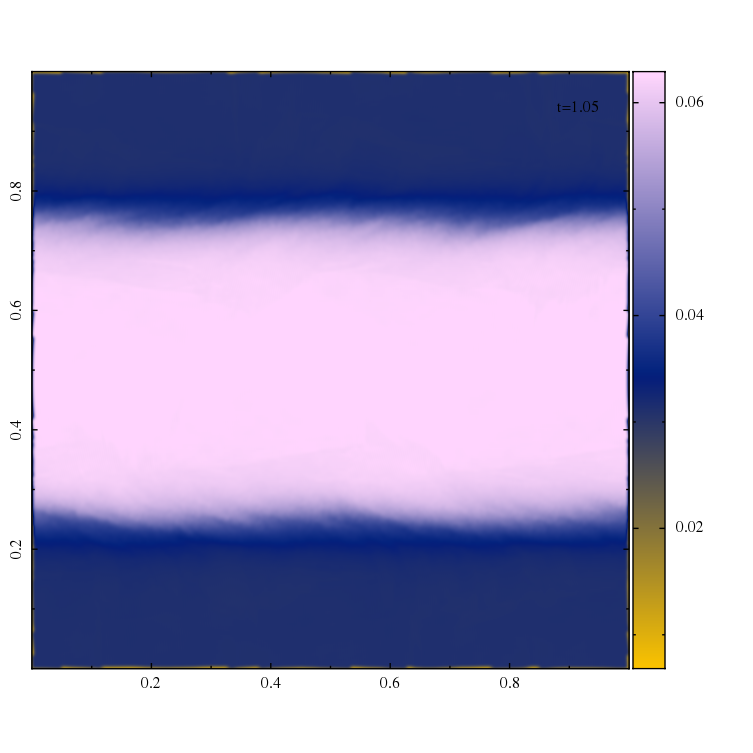} }
 \subfloat{ \includegraphics[width=0.5\textwidth]{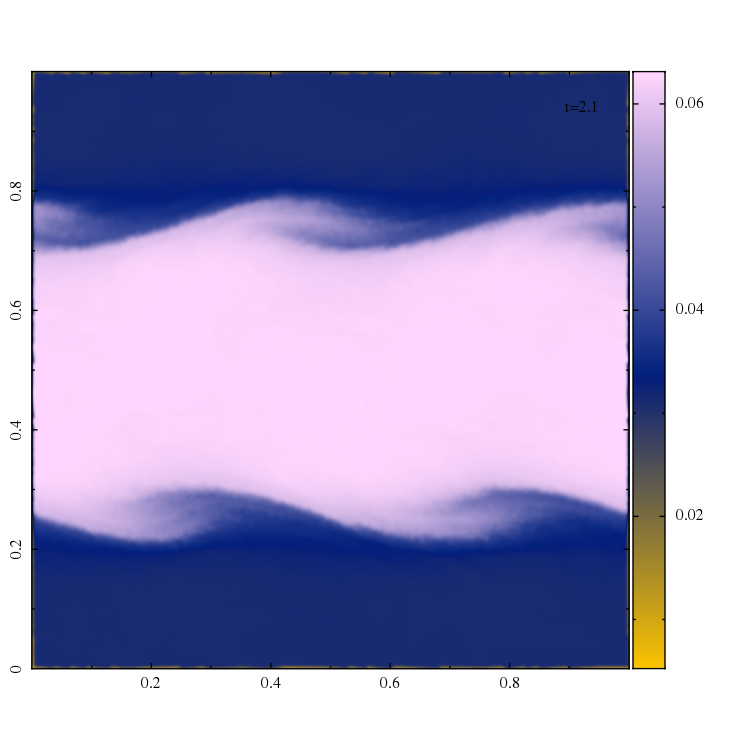} }\\
\vspace{-15pt}
 \subfloat{ \includegraphics[width=0.5\textwidth]{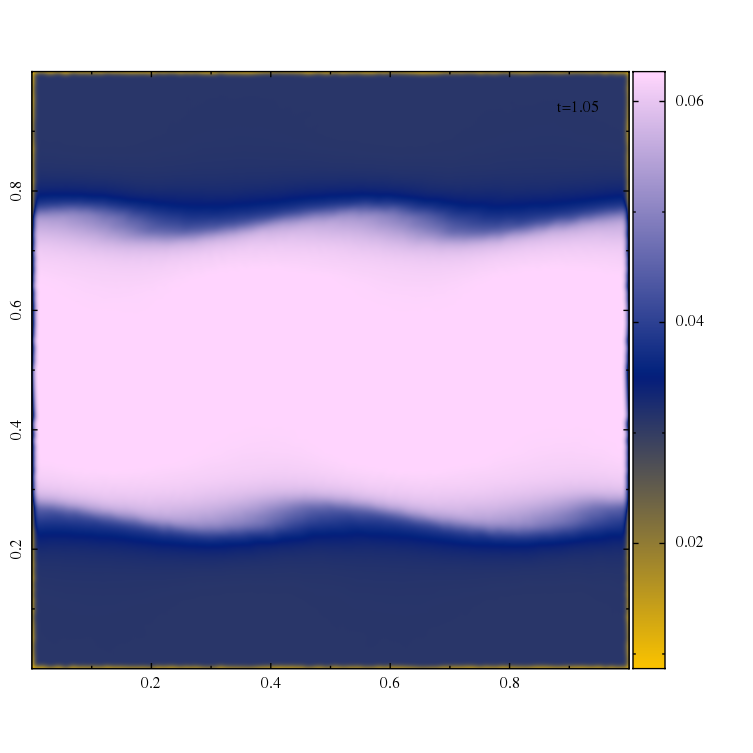} }
 \subfloat{ \includegraphics[width=0.5\textwidth]{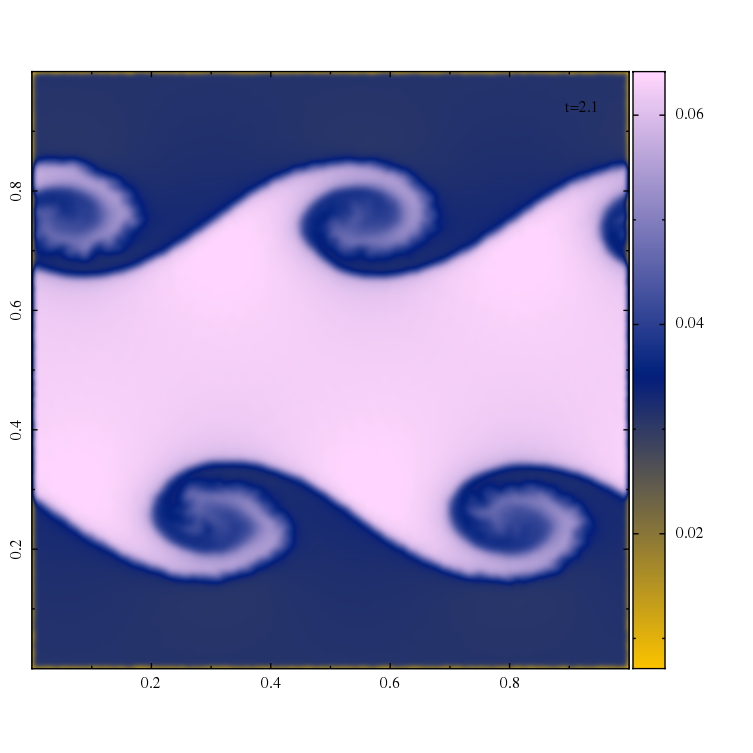} }
\vspace{-15pt}
\caption{(Continuation) Projected density plots for the simulations of set B. 
%Each column shows the perturbations at $\tau_B$ (left) and $2\tau_B$ (right). 
From top to bottom: BOSPH96 and BOSPH442.}
\end{figure*}

The mass varying initial conditions in the simulations of set
B lead to pronounced perturbations with well developed rolls being produced 
only in the runs
with a maximum number of neighbors (Fig. \ref{B}). Small undulations
observed at time $t=\tau_B$ (left column) are evolved into a KHI for
$t=2\tau_B$ (right column) in runs BGADG64 and BOSPH442.
Surprisingly, standard SPH with multimass setup do develop KHI,
contrary to previous claims. Comparing a BGADG64 density projection
at $t=2.1$ to a reference solution obtained with the {\it Pencil}
code of \citet{McNally12} (their Fig. 2) we observe a very similar
shape and amplitude of density rolls. However, the reference
solution shows the density projection at $t=1.5$, indicating that in
BGADG64 the KHI develops slowly. In the case of BOSPH442, the
density projection which matches approximately the reference
solution corresponds to $t=1.7$. In order to check the effect of a low
number of neighbors we repeated the BGADG64 with 33 neighbors (not
shown), and obtained a similar situation to BOSPH64 with no KHI. The
Mach number of the low density layer is $M_2\approx 0.34$.

In order to explore the effect of different shear velocities and
perturbation amplitudes, we performed some additional simulations with
both codes reducing the initial velocities, and either reducing or
keeping the same perturbation amplitude $\delta {\mbox v}_y$. These
models are listed in Table \ref{tab1} as set C, where the last
letter in the name stands for large amplitude (L), $\delta {\mbox
v}_y=0.01$, and for small amplitude (S), $\delta {\mbox v}_y=0.002$,
respectively. In this case the gas is still subsonic with $c_1=0.36$
and $c_2=0.25$, giving the Mach number of low density layer
$M_2\approx 0.4$. The Fig. \ref{C} compares the results obtained
with both codes. While these appear very similar, the amplitude of
the KHI looks smaller and lacks secondary ripples for the Gadget-2 code.
\begin{figure*}[ht]
\centering
 \subfloat{ \includegraphics[width=0.5\textwidth]{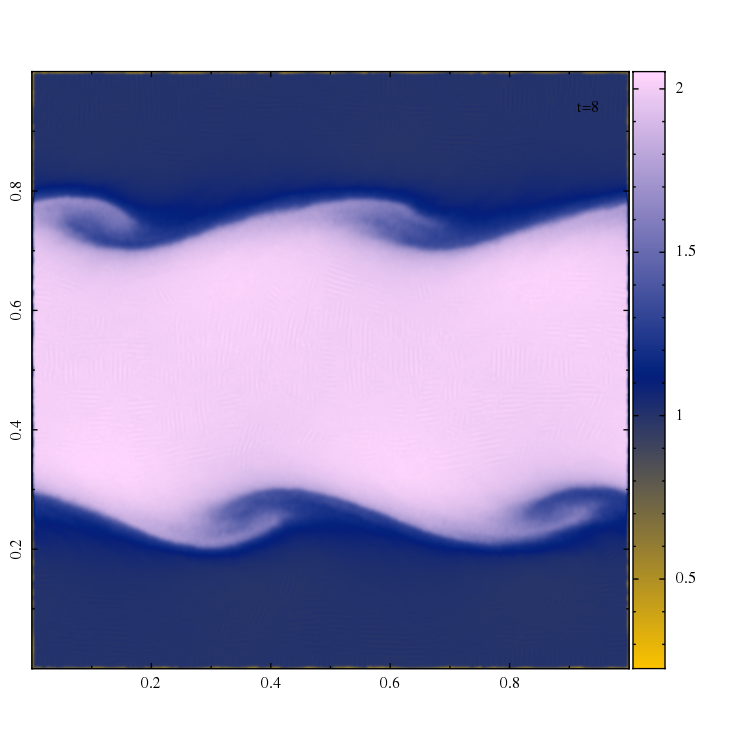} }
 \subfloat{ \includegraphics[width=0.5\textwidth]{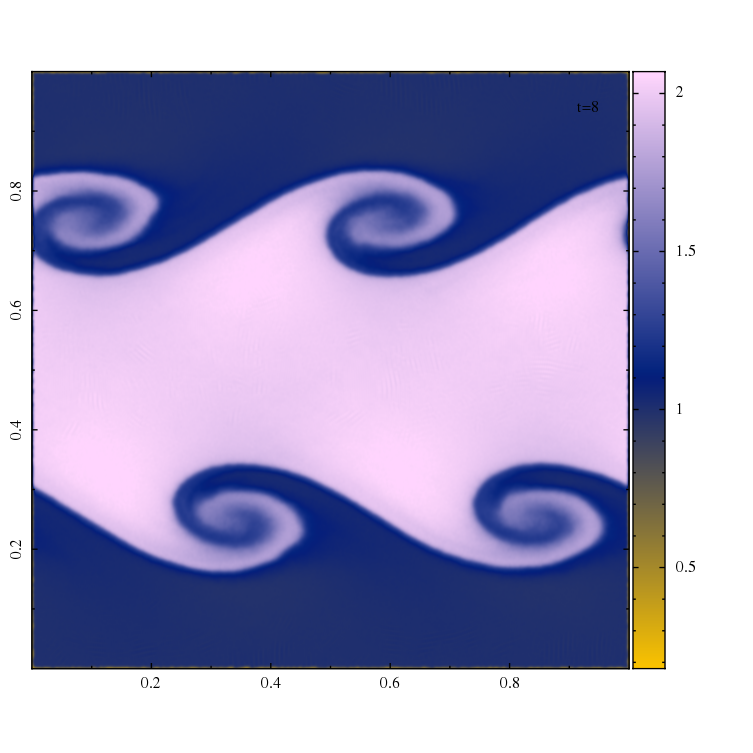} }\\
 \vspace{-15pt}
 \subfloat{ \includegraphics[width=0.5\textwidth]{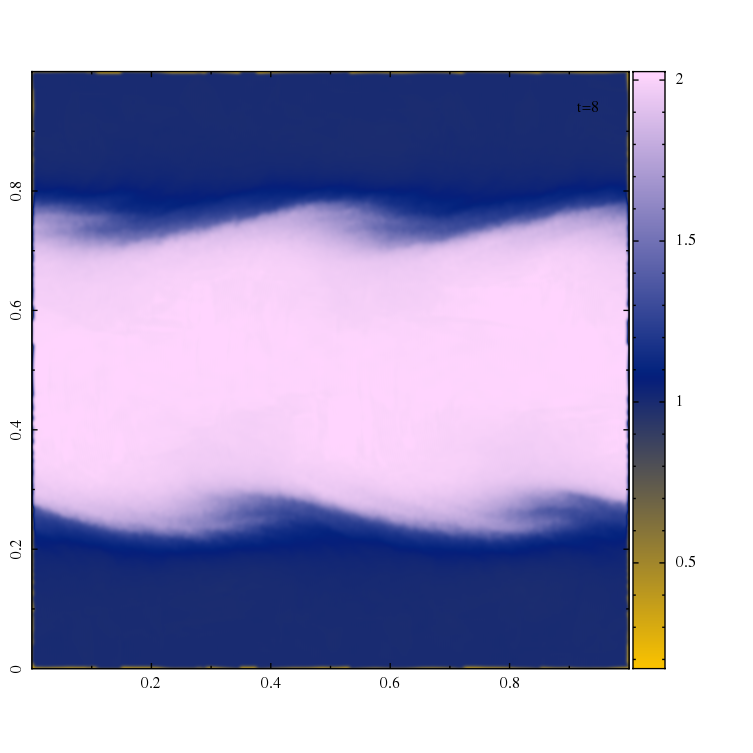} }
 \subfloat{ \includegraphics[width=0.5\textwidth]{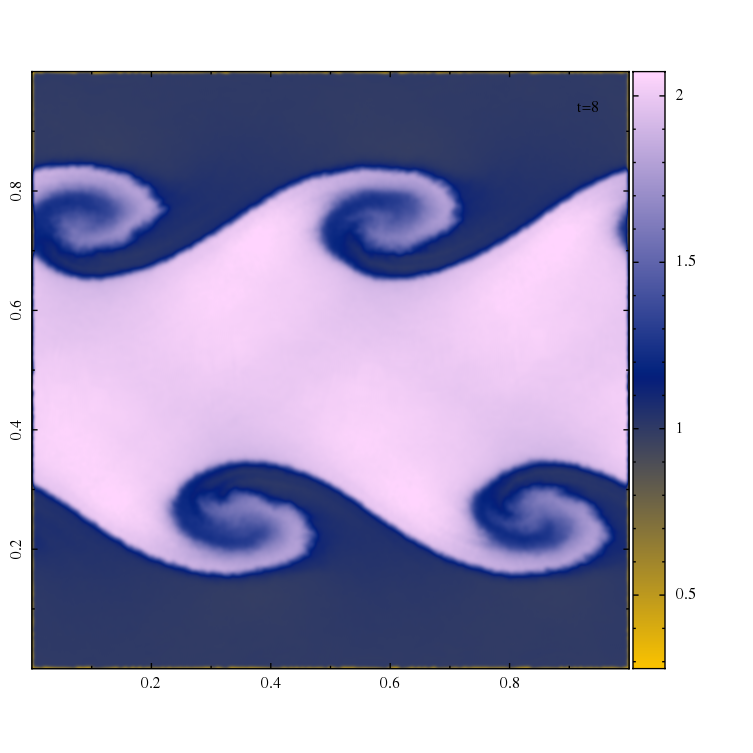} }
 \vspace{-15pt}
\caption{Projected density plots for the simulations of set C
showing the perturbations at $1.5\tau_C$. From left to right and top
to bottom: CGADG64S, CGADG64L, COSPH96S, COSPH96L.} \label{C}
\end{figure*}
%
%\FloatBarrier  %%% Barrier for figures!
%\let\clearpage\relax
\section{Concluding remarks}
\label{conc} The aim of this work was to test several implementations of
the SPH code and determine the necessary conditions for
successful simulations of the mixing process in shearing flows. We
have compared visually the projected densities for standard and optimized
SPH codes for non-linear structure formation. Runs with single mass
standard SPH formulations showed a similar behavior to that 
described in previous studies, (c.f. Fig. 13 of \citet{Agertz07} and
Fig. 7 of \citet{Price08}). On the other hand, it seems that the
multimass simulations do not depend on the SPH code implementation,
but rather on the number of neighbors. The optimized SPH code is
almost two times slower than the standard SPH for the same number of
neighbors, -- a disadvantage that was reported to be absent in the new
SPHS code of the same authors \citep{Read12}. We have shown that
for mixing problems using the SPH formalism, it is essential to take
much care in setting the initial conditions and the number of neighbors.
Further investigation of the problem is
necessary in order to obtain robust results with SPH.

\begin{acknowledgement} This work has been partially supported by ABACUS, CONACyT
grant EDOMEX-2011-C01-165873. We thank Acarus of the Universidad de Sonora, M\'exico, for the use of
their computing facilities.
\end{acknowledgement}


\begin{thebibliography}{99}

\bibitem[Agertz et al.(2007)]{Agertz07}
Agertz O., et al, (2007) Fundamental differences between SPH and grid methods. {\em MNRAS} 380, 963

\bibitem[Cha, Inutsuka \& Nayakshin(2010)]{Cha10}
Cha S.H., Inutsuka S.I., Nayakshin S., (2010) Kelvin-Helmholtz instabilities with Godunov smoothed particle
hydrodynamics. {\em MNRAS} 403, 1165

\bibitem[Dehnen \& Aly(2012)]{Dehnen12}
Dehnen W., Aly H., (2012) Improving convergence in smoothed particle hydrodynamics simulations without pairing instability. {\em MNRAS} 425, 1068

\bibitem[Hubber, Falle, \& Goodwin (2013)]{Hubber13}
Hubber D. A., Falle S. A. E. G. and Goodwin S. P., (2013) Convergence of AMR and SPH simulations - I. Hydrodynamical
resolution and convergence tests. {\em MNRAS} 432, 711

\bibitem[McNally, Lyra \& Passy(2012)]{McNally12}
McNally C.P., Lyra W., Passy J-C., (2012) A Well-posed Kelvin-Helmholtz Instability Test and Comparison. {\em ApJSS} 201, 18

\bibitem[Murray et al.(1993)]{Murray93}
Murray S. D., White S. D. M., Blondin J. M., Lin D. N. C., (1993) Dynamical instabilities in two-phase media and the minimum masses of stellar systems.
{\em ApJ} 407, 588

\bibitem[Price(2008)]{Price08}
Price D. J., (2008) Modelling discontinuities and Kelvin-Helmholts instabilities in SPH. {\em J. Comput. Phys.} 227, 10040

\bibitem[Read, Hayfield \& Agertz(2010)]{Read10}
Read J.I., Hayfield T., and Agertz O., (2010) Resolving mixing in smoothed particle hydrodynamics. {\em MNRAS} 405, 1513

\bibitem[Read \& Hayfield(2012)]{Read12}
Read J.I., Hayfield T. (2012) SPHS: smoothed particle hydrodynamics with a higher order dissipation switch. {\em MNRAS} 422, 3037

\bibitem[Robertson et al.(2010)]{Robertson10} Robertson B.E., Kravtsov A.V., Gnedin N.Y., Abel T.
and Rudd D.H. (2010) Computational Eulerian hydrodynamics and Galilean invariance. {\em MNRAS} 401, 2463

\bibitem[Springel(2005)]{Springel05}
Springel V., (2005) The cosmological simulation code GADGET-2. {\em MNRAS} 364, 1105

\bibitem[Vietri, Ferrara \& Miniati(1997)]{Vietri97}
Vietri M., Ferrara A., Miniati F., (1997) The survival of interstellar clouds against Kelvin-Helmholtz instabilities {\em ApJ} 483, 262

\end{thebibliography}
\end{document}